\documentstyle[12pt,aaspp4]{article}

\begin{document}

\def\kms{km~s$^{-1}$}
\def\msun{$M_{\odot}$}
\def\rsun{$R_{\odot}$}
\def\lsun{$L_{\odot}$}
\def\halpha{H$\alpha$}
\def\Teff{T$_{\rm eff}$}
\def\logg{$log_g$}

\tighten

\title{ 5-micron photometry of late-type dwarfs}

\author {I. Neill Reid\footnote{Visiting Astronomer at the Infrared Telescope
Facility, which is operated by the University of Hawaii under contract from the National
  Aeronautics and Space Administration.}}
\affil{Space Telescope Science Institute, 3700 San Martin Drive,
Baltimore, MD 21218 \\
\& Dept. of Physics \& Astronomy, University of Pennsylvania, 209 S. 33rd Street, 
 Philadelphia, PA 19104-6396 \\
e-mail: inr@stsci.edu}

\author {K. L. Cruz$^1$}
\affil {Dept. of Physics \& Astronomy, University of Pennsylvania, 209 S. 33rd
Street, Philadelphia, PA 19104-6396 \\
e-mail: kelle@hep.upenn.edu}

\begin {abstract}

We present narrowband-M photometry of nine low-mass dwarfs with
spectral types ranging from M2.5 to L0.5. Combining the (L$'$-M$'$) colours
derived from our observations with data from the literature, we find
colours consistent with a Rayleigh-Jeans flux distribution for spectral
types earlier than M5, but enhanced $F_{3.8} \over F_{4.7}$ flux ratios (negative
(L$'$-M$'$) colours) at later spectral types. This probably reflects 
increased absorption at M$'$ due to the CO fundamental band.
We compare our results against recent model predictions and briefly 
discuss the implications.

\end{abstract}


\section {Introduction}

Ultracool dwarfs, with spectral types later than M6.5, have effective
temperatures of less than 2700K. As a result, a significant fraction
of the total energy is emitted at infrared wavelengths, and mapping
the spectral energy distribution beyond $\lambda = 1 \mu$m is crucial to
determining accurate bolometric magnitudes. Ground-based observations
at those wavelengths have to cope with strong absorption features due to 
water and CO$_2$ in the terrestial atmosphere, but broadband 
photometry through the atmospheric windows at 1.25$\mu$m (J),
1.6$\mu$m (H), 2.2$\mu$m (K) and 3.5 $\mu$m (L) has been available
for M dwarfs for over three decades.  Recent large-scale surveys,
notably 2MASS (Skrutskie {\sl et al.}, 1997), provide extensive JHK data 
for the even cooler L and T
dwarfs. Narrowband spectrophotometry from 1 to 2.5$\mu$m, extending through terrestial
water bands, was obtained for a handful of M dwarfs by Reid \& Gilmore (1984), 
and extended to 3.5$\mu$m for a subset of those stars by Berriman \& Reid (1987).
With the development of array detectors, low-resolution near-infrared spectroscopy now
exists for a representative sample of over 40 
late-M, L and T dwarfs (Jones {\sl et al.}, 1996;
Leggett {\sl et al.}, 2000; Reid {\sl et al.}, 2001a; Burgasser {\sl et al.}, 2001; 
Geballe {\sl et al.}, 2001).

Spectral energy distributions of cool dwarfs are less well-defined at longer
wavelengths, particularly for the latest spectral types. The 4.8$\mu$m (M) and
10.2$\mu$m (N) bands are centred on lower-throughput atmospheric windows 
than the J, H, K or even L bands, and 
observations become increasingly difficult with the growth in
the thermal background. Spaceborne observations avoid this problem, but 
both IRAS and ISO had relatively low sensitivity, and provide observations
of only a few of the brightest sources, predominantly early-type dwarfs.
Ground-based observations are similarly limited. Berriman \& Reid (1987)
obtained M-band photometry of seven nearby dwarfs, including VB8 (spectral type M7), 
and also report longer-wavelength spectrophotometry (by Aitken \& Roche) of
the M5.5 dwarf Gl 406 (Wolf 359). Most recently, Leggett {\sl  et al.} (2001b -
hereinafter, LSDSS) 
have supplemented those observations with data for four ultracool dwarfs, two 
of spectral type L and two of spectral type T. 

The forthcoming SIRTF mission should provide more detailed mid-infrared spectrophotometry
of a representative sample of cool dwarfs, including isolated examples of both
L and T dwarfs. In the meantime, we have combined new ground-based 5$\mu$m photometry
of M and L dwarfs with data from the literature in an attempt to assess the likely
flux levels as a function of spectral type. The following section describes the
photometric system used to obtain the additional observations, and presents the
resultant photometry. Those data are combined with the relatively sparse set of
previous observations at mid-infrared wavelengths in Section 3, and compared against
predictions of recent theoretical models by Chabrier {\sl et al.} (2001). 
Finally, we summarise our conclusions.

\section {5-micron Photometry}

\subsection {Infrared fluxes and the M$'$ passband}

The original infrared broadband photometry system was devised by Johnson (1964)
to take advantage of the highest-transparency atmospheric windows at those wavelengths, 
and was limited to the JKLMN passbands. Johnson's original observations of
K and M dwarfs includes JKL photometry for only 17 stars (Johnson, 1965), with
the 1.6$\mu$m H band only added at a later date (Johnson {\sl et al.}, 1968).
The L and M bands in this system are both extremely broad (FWHM$\approx 0.5 \mu$m), with 
effective wavelengths of 3.45 and 4.75$\mu$m, respectively. 
Figure 1 matches the transmission curves for those two passbands (from Bessell \&
Brett, 1988) against the average atmospheric transparency at Mauna Kea Observatory.
Both filters extend significantly beyond the atmospheric windows, sampling the
3$\mu$m and 5$\mu$m H$_2$O bands and 4.3$\mu$m CO$_2$ absorption - wavelength
regions with both high opacity and high emissivity. Not only does
the mismatch lead to reduced flux from the astronomical target, but also
high background levels. 

The original L-band observations were made using PbS detectors, and their
decline in sensitivity longward of 3.5$\mu$m prevented a more favourable
centering in wavelength.
Once InSb detectors became available, that mismatch could be corrected, and
the L$'$ passband, $\lambda_{eff} = 3.80 \mu$m, was defined. 
Slightly broader than
the L-band (FWHM$\approx 0.54 \mu$m), the L$'$ band is better matched to
the atmospheric transmission curve. This passband has been the standard for
broadband 3.5$\mu$m photometry since the mid-1980s, albeit with subtle
variations between individual systems, as discussed further in Section 3. 

The situation at 5$\mu$m is more complicated, since terrestrial absorption
is present to some extent over the full wavelength range. However, it is
clear that the Johnson passband can be improved. At the IRTF, this goal has
been achieved by defining a narrowband-M system, which we denote here as M$'$, with
$\lambda_{eff} \sim 4.68\mu$m and FWHM$\approx 0.24\mu$m. As Figure 1 shows,
the passband is centred in the optimum region of the 5$\mu$m window. The
observations discussed below were made using this filter, which is part
of the MKO-NIR system (Simons \& Tokunaga, 2001; Tokunaga \& Simons, 2001). 

At present, there are no extensive observations of photometric standards in
the M$'$ passband. Our photometry is therefore calibrated through observations
of M-band standards from Sinton \& Tittemore (1984: S\&T), 
specifically $\zeta^2$ Ceti, $\eta$ Vir.
$\rho$ Vir and $\tau$ Her. All four are early-type stars (B9 III, A2 IV, A0 V and A), and
are therefore likely to have spectral energy distributions closely matching
the Rayleigh-Jeans approximation at these wavelengths. In that case, 
the flux zeropoint for the M$'$ system 
is given by scaling the M-band zeropoint by $({\lambda_{M'} \over \lambda_M})^{-4}$.
Bessell \& Brett (1988) cite a zeropoint of 
F$_\lambda(0) = 2.04 \times 10^{-11} \ {\rm W \ m^{-2} \mu m^{-1}}$ for the M band,
giving F$_\lambda(0) = 2.13 \times 10^{-11} \ {\rm W \ m^{-2} \mu m^{-1}}$, or 
F$_\nu(0) = 156$ Janskys at M$'$.

As a consistency check, we observed both S\&T standards and
stars from the UKIRT standard list (specifically HD 105601, HF 136754 and Gl 811.1)
in May 2001. In addition, all of the stars observed during that run (including the
M dwarfs) were measured
using both M and M$'$ filters. Our results show marginal evidence for a $0.06\pm0.03$
magnitude offset between the S\&T and UKIRT systems, with the UKIRT system giving
brighter magnitudes for the programme stars. There is no evidence, however, for a 
systematic difference in the relative magnitudes at M and M$'$ between either set
of standards and the target stars; that is, there is no evidence for a colour term in
(M-M$'$) for spectral types between A0 and M5.5 (Gl 406).
  
Finally, have allowed for atmospheric extinction using values of 0.09 mag/airmass at
L$'$ (Krisciunas {\sl et al.}, 1987) and 0.19 mag/airmass at M$'$. The latter
value is cited on the UKIRT web-site as appropriate for a narrowband-M filter.
Since all of our observations, both of standards and
program objects, were taken at airmasses of less than 1.30, the corrections are
less than $\pm0.03$ magnitudes.

\subsection {Observations}

Our observations were made on January 19, 20 and May 14 (UT) 2001. We used 
NSFCam, an imaging camera equipped with a $256 \times 256$ InSb array, 
on the NASA Infrared Telescope Facility. Conditions were excellent
on all three nights, with photometric skies, extremely low water-vapour content and 
seeing of 0.4 to 0.6 arcseconds at 3.5$\mu$m. Table 1 lists relevant data for
our targets, which include several well-known nearby stars and three bright,
ultracool dwarfs from the 2MASS-selected sample analysed by Gizis {\sl et al.} (2000).
We note that HST observations of the L dwarf 2M0746 resolve it as a binary,
separation 0.22 arcseconds (Reid {\sl et al.}, 2001b), and therefore unresolved in
our observations. The components have very similar luminosities, 
$\Delta I = 0.62$ magnitudes, so the 3 to 5$\mu$m colours should be characteristic
of spectral class L0/L1 (see Figure 5, Reid {\sl et al.}, 2001b). 

Photometry was obtained at  
L$'$ and M$'$ in our January observations, and, as discussed above, at M$'$ and
M on May 14. In both cases, we used a five-point dither pattern, centred on
the target, offsetting by $\pm3$ arcseconds in $\alpha$ and $\delta$. 
We used a pixel scale of 0.055 arcsec pix$^{-1}$
to cope with the bright sky background;
even so, maximum integration times in a single frame were limited to 
0.2-0.3 seconds, and we co-added 20 to 150 frames at each position.
The total integration times in M$'$ are 500-600 seconds for the
brighter targets (Wolf 359, LHS 292), and 2400 to 3000 seconds for the
fainter sources. 

Data reduction was undertaken in two stages. First, IDL and {\sl iraf} routines
were used to 
combine the five exposures in each dither pattern to give
a composite image.  In doing so, we compensate for the sky background
by subtracting exposures adjacent in time; measure the centroid
of the target; and use shift-and-add techniques to align the five
exposures. Finally, all composite images of a
given target are summed to give the final combined image. 
This technique requires that either the target 
or a nearby source is visible on each individual exposure.

Once the individual frames are sky-subtracted and combined, we 
used the aperture photometry routine {\sl phot} in the {\sl iraf} package to
determine instrumental magnitudes for each target. Based on the 
curve of growth, we adopt an aperture of radius 26 pixels (1.43 arcseconds),
measuring the sky background at $1.65 < r < 2.09$ arcseconds. Most of
our targets have previously published L$'$ data, but 2M0027, 2M1441 and
2M0746 lack such data. We have used our observations to derive L$'$ for those
stars. The data are calibrated against our measurements of Gl 406 (Wolf 359), adopting 
Leggett's (1992) photometry for the M5.5 dwarf. We also obtained L' data for LHS 2065, 
and derive L$'$=9.43, as compared with L$'$=9.19 given by Leggett (1992). The latter
value gives (K-L$'$)=0.75, substantially redder than other M8 dwarfs observed here and by 
LSDSS. We therefore adopt our L$'$ measurement for this star.

As noted above, the photometric zeropoint at M$'$ is set by observations of 
standards from Sinton \& Tittemore (1984); the zeropoints determined from
individual standards agree to $\pm0.02$ magnitudes. Table 2 lists the
photometry for our targets. We include data for the M2.5V UKIRT standard, Gl 811.1, 
where the M$'$ data are calibrated against our measurements of the S\&T standards
(UKIRT lists M=6.73). 
Table 2 also lists M$'$-band bolometric corrections, derived by combining
our photometry with either previously-measured corrections at other wavelengths, 
or by assuming BC$_J$=1.9 magnitudes (see Reid {\sl et al.}, 2001a,b). 
The only stars with previous 5$\mu$m observations are Gl 643 and Gl 752A,
observed by Berriman \& Reid (1987), with M=6.4$\pm$0.1 and M=4.5$\pm$0.1
respectively. Those  measurements are consistent with the present photometry. 

\section {Discussion}

\subsection {Empirical results: the (L$'$-M$')$/spectral-type relation}

The main result from our analysis is illustrated in Figure 2, which plots (L$'$-M$'$)
colour as a function of spectral type. In addition to the photometry listed in Table 2, we
have plotted Berriman \& Reid's (1987) (L$'$-M) data for Gl 884 (K5), Gl 699 (M5), Gl 866
(M5.5) and Gl 644C (M7), together with the LSDSS  photometry of
2M0036 (L3.5), SD0857+57 (L8), SD1254-01 (T2) and 2M0559-14 (T5). Our L$'$ photometry
and the Leggett {\sl et al.} data are on the MKO-NIR system, while the Berriman \&
Reid measurements are on the UKIRT system. As yet, there is no extensive
published comparison between these systems, but the available data indicate that
systematics differences are relatively small for K and M dwarfs.
Leggett (1992) and Leggett {\sl et al.} (1998) list UKIRT-system L$'$ photometry 
for seven stars (spectral types M4 to M9) in common with LSDSS. 
A comparison gives
\begin{displaymath}
L'_{MKO} \ - \ L'_{UKIRT} \ = \ 0.013 \pm 0.050
\end{displaymath}
with no statistically significant trend against spectral type 
(the two largest residuals are -0.07 mag. for
LHS 2397a, M8,  and +0.10 mag. for LHS 2924, M9). 
Thus, the L$'$ photometry of M dwarfs can be regarded as self-consistent 
to better than $\pm5$\% between the UKIRT and MKO-NIR systems.

At the longer wavelength, our own observations show that the
Berriman \& Reid M-band data should be compatible with the M$'$ system. In any case, 
the data considered here (17 objects) constitute the totality of 
M-band photometry currently available 
for late-type dwarfs. Spectral types for all of the Berriman \& Reid Gliese stars
are taken from Reid {\sl et al.} (1995), the value for
2M0036 (L3.5) is from Kirkpatrick {\sl et al.} (1999), while the remaining types are given
in LSDSS. 

Figure 2 also plots (K-L$'$) data for the M-band sample and other late-type dwarfs.
The latter data are taken from Leggett (1992 - spectral types K5 to M9), 
Leggett {\sl et al.} (1998 - M0 to M6.5) and LSDSS (M6 to T8).
As noted above, the L$'$ data in the first two references are on the UKIRT system,
but, since observations are limited to K and M dwarfs, the entire dataset can be
regarded as consistent  with the MKO-NIR system to $\pm5$\%. 

While the parameter space remains sparsely sampled and uncertainties are still 
significant, the data show a clear trend with spectral type. Earlier than M4, the
colours are broadly consistent with (L$'$-M$') \approx 0$, indicating that the
flux ratio, $F_{L'/M'} = {F_{3.8} \over F_{4.7}}$, is close to that expected for 
a Rayleigh-Jeans spectral energy distribution, F$_\lambda \propto \lambda^{-4}$.
The increasing complexity of molecular absorption in later-type dwarfs leads to
significant departures from blackbody distributions with decreasing temperature, 
so it is scarcely surprising that later-type dwarfs show increasing
deviation from this ratio. The (L$'$-M$'$) colours become negative at 
spectral type M5/M6, indicating  proportionately higher flux in the 
shorter wavelength band. This transition lies close to the temperature where dust is 
predicted to form in sufficient quantities to influences M dwarf atmospheres. As 
 originally pointed out by Tsuji {\sl et al.} (1996), dust can modify the
energy distribution, notably the water bands between 1 and 3 $\mu$m, by
warming the outer atmospheric layers. 
However, the absence of a corresponding feature in (K-L$'$)
suggests that the change in (L$'$-M$'$) reflects reduced flux 
at 4.7$\mu$m, rather than  enhanced flux at 3.8$\mu$m. 

A possible candidate for the observed behaviour is increased CO absorption in the 
5$\mu$m fundamental band. The existence of this feature at these spectral types
is no surprise, since the 2.2$\mu$m overtone band is evident in dwarfs as early as K5 (Reid 
\& Gilmore, 1984), while Noll {\sl et al.} (1997) detect
broad CO in their spectrum of Gl 229A (M1.5). Thus, the feature should be 
well developed by spectral type M5. It is not clear, however, whether CO 
alone is responsible for the change in (L$'$-M$'$) colour. Unfortunately,
Gl 229A is the only M dwarf with useful spectroscopic observations 
at these wavelengths. 
The few M dwarf spectra obtained by ISO (Wehrse {\sl et al.}, 1997;
Tsuji {\sl et al.}, 1997) have low signal-to-noise and uncertain calibration, 
and set weak constraints on the flux distribution from 4 to 6$\mu$m. 

\subsection {A comparison with models: KL$'$M$'$ colours}

We can compare the broadband observations against predicted colours derived from
the latest set of theoretical models by Chabrier {\sl et al.} (2000). Those `DUSTY'
models take account of dust species in the calculation of both the equation of state and 
opacity, besides allowing for scattering and absorption by dust in determining
the radiative transfer equation. Chabrier {\sl et al.} present predicted VRIJKL$'$M 
absolute magnitudes for low mass ($M \le 0.1 M_\odot$) dwarfs at ages from 0.1 to 10 Gyrs.

The L$'$ magnitudes synthesised from the DUSTY models are 
effectively on the UKIRT system, while the M-band magnitudes are matched to the
Johnson system. 
As discussed in the previous section, this leads to only small
differences in the M- and early L-dwarf r\'egimes
with respect to the observations collected in Figure 2.  However, there
are more significant differences at later spectral types. LSDSS
estimate that M and M$'$ measurements differ by 4\% at 2000K (L0) and by
9\% at 950K (T5), with M fainter in both cases. At L$'$, synthesised
magnitudes from theoretical models suggest that the UKIRT system
gives brighter magnitudes than the MKO-NIR system by $<$0.05 mag. at L0,
$\sim0.1$ magnitude at L8 and $\sim0.2$ magnitudes for T dwarfs 
(Stephens {\sl et al.}, 2001).
These corrections are not yet determined precisely, so we have not attempted to
adjust the models to match the MKO-NIR system. 
In addition, the models are known to treat dust incorrectly in T dwarfs, leading
to infrared colours which are too red. The latter issue is discussed in more detail
by LSDSS; our main concern are the M  and early L dwarfs. 
 
The models predict colours as a function of effective temperature, and 
comparison with observations therefore
requires the adoption of a temperature/spectral type relation. That relation
remains uncertain by $\pm150$K for late-type dwarfs. Our M dwarf scale is
tied to Leggett {\sl et al.'s} (1996) observations of GJ 1111 (M6.5) at T$_{eff}=2700$K,
while current concensus places the boundary between the M and L spectral types at 
T$_{eff} = 2050\pm100$K (Basri {\sl et al.}, 2000; Kirkpatrick {\sl  et al.}, 2000;
Schweizer {\sl et al.}, 2001). 
The L dwarf scale follows the scheme outlined by Reid {\sl et al.} (1999), with the L/T 
transition at 1300/1400 K and Gl 229B (T6 - Geballe {\sl et al.} 2001) at $\sim950$K. 

Figure 2 shows the predicted colours for 1 and 10 Gyr. DUSTY models. At
the infrared wavelengths considered here, those colours are primarily temperature
dependent.
Given the uncertainties, the model predictions are in good agreement with the 
observed colours in the (L$'$-M$'$) plane. Following the discussion above, the
models are likely to be systematically too blue by $\sim0.1$ magnitude at L0
(-0.05 mag. at L$'$; +0.04 at M) and $\sim0.15$ mag. too blue at L8 (-0.1 mag.
at L$'$; +0.05 mag. at M). These corrections move the predicted relations close 
to (L$'$-M$'$)=0 at spectral types $\approx$L3 to L8, 0.2 to 0.3 magnitudes 
(2$\sigma$) redder than 
the observations. It is not clear whether the
models match the observed transition to Rayleigh-Jeans flux ratios at earlier
spectral types than M5, since the coolest model listed by Chabrier {\sl et al.} has
T$_{eff}=2971$K (spectral type $\approx$M4.5). 

In the (K-L$'$) plane the agreement between models and observations 
is less precise, although systematic differences between the UKIRT and MKO-NIR
systems go some way to reducing the discrepancies. As noted above, the
substantial offset between theory and observation for T dwarfs is not
unexpected, and is considered in more detail by
LSDSS. There are lesser discrepancies at earlier
types, notably between M8 and $\sim$L5/L6, where the models can be 0.3 magnitudes
magnitudes redder than the observations, although systemic differences probably
reduce the discrepancy to 0.25 magnitudes. 

\subsection {A comparison with models: colour-magnitude diagrams}

These results have potentially significant implications for analysis
of infrared surveys, such as the proposed SIRTF Legacy projects, for 
very low-mass stars and brown dwarfs. Theoretical models have identified the 5$\mu$m 
spectral region as a prime passband for searching for these low temperature
objects. However, combining the (L$'$-M$'$) and (K-L$'$) colours from our
observations and the LSDSS measurements indicates that the
(K-M$'$) colours of late-M and L dwarfs are bluer than expected. Figure 3 illustrates
this  plotting the (M$_M$, (K-M$'$)) diagram for the M and L 
dwarfs in Table 2 and for 2M0036+18 from LSDSS sample. The other
three dwarfs with M$'$ photometry in the latter sample lack distance measurements
at present.
The absolute magnitude for 2M0746 has been corrected by +0.7 magnitudes to allow
for its binarity.
In comparison, we plot 0.5-, 1- and 5-Gyr isochrones from Chabrier {\sl et al.} (2000). 
While the M dwarfs lie within 0.1 to 0.2 magnitudes (in (K-M)) of the isochrones,
the L dwarfs show larger offsets, with, depending on its age, 2M0036+18 lying 0.35 to 
0.6 magnitudes blueward of the theoretical isochrones.
 
There are at least three dimensions to the comparison made in
figure 3: colour, absolute magnitude and age. An isochrone comparison
tends to focus on colour/age at the expense of absolute magnitude. It is
therefore instructive to make direct comparison between the models and
empirical data for individual objects. 
The uncertainties in the temperature/spectral-type relation complicate such
comparisons, but we consider two test cases:
\begin{description}
\item{\bf 2M0746} is the less controversial example. 
With a spectral type of L0.5, most temperature calibration
schemes assign a temperature of $\approx2000$K (see, for example, Schweizer 
{\sl et al.}, 2001), while the dwarf has a well-determined parallax and has been
observed over a wide wavelength range. There is no detectable lithium
absorption, setting a lower limit of 0.06M$_\odot$. 
Table 3 compares some of its
observed properties (allowing for binarity) 
against DUSTY predictions for dwarfs of similar temperature, 
spanning a range of ages. The model (K-M) colours are 0.28 to 0.38 
magnitudes redder than 2M0746 (0.24 to 0.34 magnitudes, allowing for the 
likely M/M$'$ systemic
offset). Moreover, while the predicted M-band absolute magnitudes lie within 
0.2 magnitudes of the observed datum (hence our choice of M$_M$ in Figure 3), 
2M0746 is 0.23 to 0.58 magnitudes brighter than the predictions at M$_K$. 
Table 3 also compares the observed and predicted (I-K) colours; the models are
0.5 to 0.8 magnitudes redder than the data.

\item{\bf 2M0036} is more problematic, since with a spectral type of L3.5, it
falls in the region where temperature scales start to diverge.
Schweizer {\sl et al.} (2001), however, derive a temperature estimate of 1800K, based 
on Keck HIRES data, and we adopt that value for present purposes. As with 2M0746,
there is no lithium absorption, excluding the 0.5-Gyr. model as a potential match.
There is no suitable 1 Gyr model in the suite presented by Chabrier {\sl et al.},
but the 0.072$M_\odot$ 5 and 10 Gyr models both predict M-band magnitudes
in good agreement with the observations. Again, the K-band absolute magnitudes
are fainter than the observed value  by $\sim0.4$ magnitudes, the predicted
(I-K) colours are up to 0.8 magnitudes redder than the observations, and the
theoretical
bolometric magnitudes are substantially fainter than the measured value\footnote{
Chabrier {\sl et al.} note that the observed (R-I) colour for 2M0036 is significantly
bluer than the DUSTY isochrones, and suggest that this stems from photometric error.
We prefer the more traditional alternative of modifying the model to match the data}. 
\end{description}
Dust has become the universal panacea for reconciling theory and observations of
cool dwarfs, but while variations in dust content might explain the scatter in
(J-K) colours observed as a function of spectral type (Kirkpatrick {\sl et al.}, 1999;
LSDSS), the optical/infrared colours suggest a more complex origin
for the discrepancies noted here. If backwarming were responsible for the higher
K-band fluxes observed in both L dwarfs, one might expect higher blanketting (than
in the models) at 
optical wavelengths, and redder optical/infrared colours than predicted. In fact,
Table 3 shows that the colours differ in the {\sl opposite} sense. We note that
the H$_2$O line lists used in the computation of the DUSTY models are known 
to be incomplete at near-infrared wavelengths (Chabrier, priv. comm., 2001), and
this undoubtedly contributes to the K-band discrepancies. 

Figure 3 and Table 3 clearly indicate the importance of using 
empirical measurements at mid-infrared wavelengths to test current
calibration of theoretical models of ultracool dwarfs. Those
 models will be extremely important
in interpreting survey data obtained  by SIRTF's Infra-Red Array Camera (IRAC) 
for the purpose of determining the local space density of ultracool dwarfs. 
If the model fluxes are overestimated (by, for example, applying model
(K-M$'$) colours to empirical K-band measurements), then distances to observed sources 
will be overestimated, leading to 
an underestimate of the local number density. The present comparison suggests that,
in the case of the DUSTY models, using the theoretical M-band predictions
directly to predict number counts is more reliable than following methods which
scale fluxes relative to the currently better-observed near-infrared passbands.
SIRTF will itself acquire additional calibration data through broadband observations
of known nearby low-mass stars and brown dwarfs, but additional, higher spectral resolution
ground-based data would be extremely useful in tracing the detailed spectral
evolution at these wavelengths.

\section {Summary and conclusions}

We have presented 5$\mu$m photometry for a sample of nine dwarfs with spectral types from 
M2.5 to L0.5. Combined with similar data from the literature, these observations show
that $\langle({\rm L}' - {\rm M}') \rangle \sim 0$ for early-type M dwarfs and
$\langle({\rm L}' - {\rm M}') \rangle \sim -0.3 $ for late-type M and L dwarfs, with
a relatively sharp transition at spectral types M5/M6. 
The lack of a corresponding feature in (K-L$'$) suggests that enhanced absorption at M$'$ is
responsible, and the change in colour may be linked to increased  CO absorption at 5$\mu$m.
Additional 3 to 5$\mu$m spectroscopic observations of mid-type M dwarfs, comparable 
to those obtained
of Gl 229A by Noll {\sl et al.} (1997),  would be particularly
useful in confirming the origin of this behaviour and mapping the overall 
spectral evolution.
 
We have compared our results against theoretical predictions of the variation in
colour with spectral type, derived from the 
DUSTY models of Chabier {\sl et al.} (2000). The agreement between theory and
observation is good in (L$'$-M$'$), although the models do not extend to 
temperatures above the observed colour change. At (K-L$'$), the DUSTY models are
up to 0.25 magnitudes redder than the data between spectral 
types M7 to L4, once due allowance is made for the different photometric
systems. Matching the observations against isochrones in the (M$_M$, (K-M)) plane
shows that the models are $>0.2$ magnitudes redder than the observations.
Direct comparison against empirical data for two bright L dwarfs suggests that
this results primarily from an underestimate of the K-band flux, probably
due to incomplete water line lists in the models. 
These issues need to be taken into account
in deriving space densities of very low-mass stars and brown dwarfs from analysis of
large-area, mid-infrared surveys, such as those scheduled for the SIRTF mission.

\acknowledgments { 
We thank the referee, Sandy Leggett, for useful comments which improved the 
paper substantially and led to its doubling in length.
This research was supported partially by a  NASA/JPL grant to 2MASS Core Project Science.
KLC acknowledges support by a National Science Foundation Graduate Research Fellowship.
This publication makes use of data products from the Two Micron All Sky Survey, which is
a joint project of the University of Massachusetts and the Infrared Processing and Analysis 
Center/California Institute of Technology, funded by the National 
Aeronautics and Space Administration and the National Science Foundation.
We thank Paul Fukumura and Dave Griep for assistance with the observations.
}

\newpage

\begin{table}
\begin{center}
{\bf Table 1: Targets}
\begin{tabular}{lrrccrrc}
\tableline\tableline
Name &  RA (J2000) & Dec & Sp. & $\pi$ (mas)& Src. &  M$_V$ & Date Obs. \\
\tableline
Gl 752A & 19 \ 16 \ 55.3 & +5 \ 10 \ 8.1 & M2.5 & $170.3\pm1.4$ &1& 10.28 & 14-5-01 \\
Gl 811.1 & 20 \ 56 \ 46.6 & -10 \ 26 \ 54.6 & M2.5 & $67.7\pm10.6$ &1& 10.65 &14-5-01 \\
Gl 643 & 16 \ 55 \ 25.2 & -8 \ 19 \ 21.3 &M3.5 & $154.0\pm4.0$ &1& 12.70 & 14-5-01 \\
Gl 406 & 10 \ 56 \ 28.7& +7 \ 1 \ 37 & M5.5 & $425\pm7$ &2& 16.59 & 19-1-01, 14-5-01 \\
LHS 292 & 10 \ 48 \ 12.5& -11\ 20 \ 8 & M6.5 & $221\pm4$ &2& 17.32 & 20-1-01, 14-5-01 \\
LHS 2065 & 8 \ 53 \ 36.0 & -3 \ 29 \ 32& M8 & $118\pm2$ &2& 19.16 & 19-1-01, 19-1-01 \\
2M0027 & 0 \ 27 \ 55.9& +22 \ 19 \ 33 & M8 & $120\pm20$ &3& & 20-1-01 \\
2M1444 &  14 \ 44 \ 17.1 & +30 \ 2 \ 14 & M8 & $80\pm16$  &3& & 19-1-01 \\
2M0746 & 7 \ 46 \ 42.5& +20\ 0 \ 32 & L0.5 & $83\pm2$ &4& 19.47 & 20-1-01 \\
\tableline\tableline
\end{tabular}
\end{center}
Spectral types are from Reid {\sl et al.} (1995) for Gliese and LHS stars, and
from Gizis {\sl et al.} (2000) for the three 2MASS dwarfs. \\ 
Absolute visual magnitudes are from Reid {\sl et al.} (1995) except for 2M0746, where the
measurement is by Dahn {\sl et al.} (2000). 
2M0746 is a near-equal luminosity binary, separation 0.22 arcseconds (Reid {\sl et al.},
2001b). \\
References for parallax data: \\
1. Hipparcos catalogue, ESA (1997); \\
2. USNO, Monet {\sl  et al.} (1992); \\
3. Photometric parallax, Gizis {\sl et al.} (2000); \\
4. USNO, Reid {\sl et al.} (2001b). \\
The final column lists the date of our L$'$ and M$'$ observations. 
\end{table}
\clearpage

\begin{table}
\begin{center}
{\bf Table 2: Photometry}
\begin{tabular}{lccccrcccc}
\tableline\tableline
Name & Sp. & (J-K) &K  &  (K - L') & (L$'$-M$'$) & M$'$ & ref. & BC$_M$ & ref. \\
\tableline
Gl 752A & M2.5 & 0.84 & 4.66 & 0.20 & $0.05\pm0.03$ & $4.41\pm0.02$ & 1,2 & 2.9 & 6 \\
Gl 811.1 & M2.5 & 0.83 & 6.93 & 0.24 & $0.01\pm0.03$ & $6.68\pm0.01$ & 1, 3 & 2.9 & 6 \\ 
Gl 643 & M3.5 & 0.80 & 6.74 & 0.30 & $0.09\pm0.03$ & $6.35\pm0.02$ & 1, 2 &3.1 & 6 \\ 
Gl 406 & M5.5 & 0.98& 6.08  & 0.39  &  $-0.16\pm0.04$ & $5.85\pm0.03$ & 1, 2 & 3.2 & 6\\
LHS 292 & M6.5 & 0.94 & 7.96 &0.53  & $-0.22\pm0.07$ & 7.65$\pm0.05$ &  1, 2 & 3.25 & 6\\
LHS 2065 &  M8 & 1.26 & 9.98 &0.59 &$-0.23\pm0.10$ & 9.62$\pm0.07$ &1, 4 & 3.5 & 7\\
2M0027 & M8 & 1.05 & 9.56 & 0.41 &$-0.24\pm0.12$ & 9.39$\pm0.1$ & 4, 5 & 3.1 & 8\\
2M1444 &  M8 & 1.11 & 10.57 & 0.48 & $-0.30\pm0.12$ & 10.39$\pm0.1$ & 4, 5 & 3.2 & 8\\
2M0746 &  L0.5 & 1.24& 10.49 & 0.80 & $-0.33\pm0.10$& 10.02$\pm0.07$ & 4, 5 & 3.7 & 9\\
\tableline\tableline
\end{tabular}
\end{center}
References: Photometry\\
1. JK from Leggett, 1992 (CIT system); 
2. L$'$ from Leggett, 1992 (UKIRT system);  \\
3. L$'$, M$'$ from UKIRT photometric standard star list; \\
4. L$'$ from our observations (MKO system); 
5. JK$_s$ from 2MASS (see Carpenter, 2001). \\
Bolometric corrections: \\
6. m$_{bol}$ from Leggett {\sl et al.} (2000) \\
7. m$_{bol}$ from Leggett {\sl et al.} (2001a) \\
8. m$_{bol}$ derived assuming BC$_J$=1.9 (Figure 5, Reid {\sl et al.}, 2001a) \\
9. m$_{bol}$ from Reid {\sl et al.} (2001a). 
\end{table}
\clearpage

\begin{table}
\begin{center}
{\bf Table 3: Two specific comparisons}
\begin{tabular}{rrrrrrrrr}
\tableline\tableline
Source & T$_{eff}$ & Mass & M$_K$ & M$_M$ & (K-M) & M$_{bol}$ & BC$_{M}$ & (I-K)  \\
  & K & $M_\odot$ & \\
\tableline
Empirical \\
2M0746 & $\approx$2000 &$>0.06$ & 10.77 & 10.30 & 0.47 & 14.00 & 3.70 & 3.94 \\
Models \\
0.5 Gyr. & 2048 &0.06&11.00 & 10.25 & 0.75 & 14.20 & 3.95 & 4.46\\
1 Gyr. & 2012 & 0.07  & 11.24 &10.43 & 0.81 & 14.45 & 4.02& 4.71\\
5 Gyr. & 1998 & 0.075& 11.35 & 10.50 & 0.85 & 14.55 & 4.05& 4.79\\
10 Gyr. & 1998 & 0.075 & 11.35 &10.50 & 0.85 & 14.55 & 4.05& 4.79\\
Empirical \\
2M0036 & $\approx$1800 &$>0.06$ & 11.31 & 10.63 & 0.68 & 14.45 & 3.82 & 5.05 \\
Models \\
0.5 Gyr. & 1751 & 0.05& 11.35 & 10.36 & 0.99 & 14.90 & 4.24 & 5.75 \\
5 Gyr. & 1754 & 0.072 &11.68 & 10.66 & 1.02 & 15.20 & 4.54 & 5.90 \\
10 Gyr. & 1744 & 0.072 & 11.70 & 10.67 & 1.03 & 15.23 & 4.56 & 5.99 \\
\tableline\tableline
\end{tabular}
\end{center}
2M0746 - empirical data from Dahn {\sl et al.} (2000) and 
Reid {\sl et al.} (2001a), corrected for duplicity \\
2M0036 - empirical data from Dahn {\sl et al.} (2000), LSDSS and Reid {\sl et al.} (2000) \\
Theoretical data from Chabrier {\sl et al.} (2000). There is no suitable
1 Gyr model to match against 2M0036. M$_{bol}$ is calculated assuming M$_{bol}(\odot) = 4.75$.
\end{table}
\clearpage
\begin{figure}
\plotone{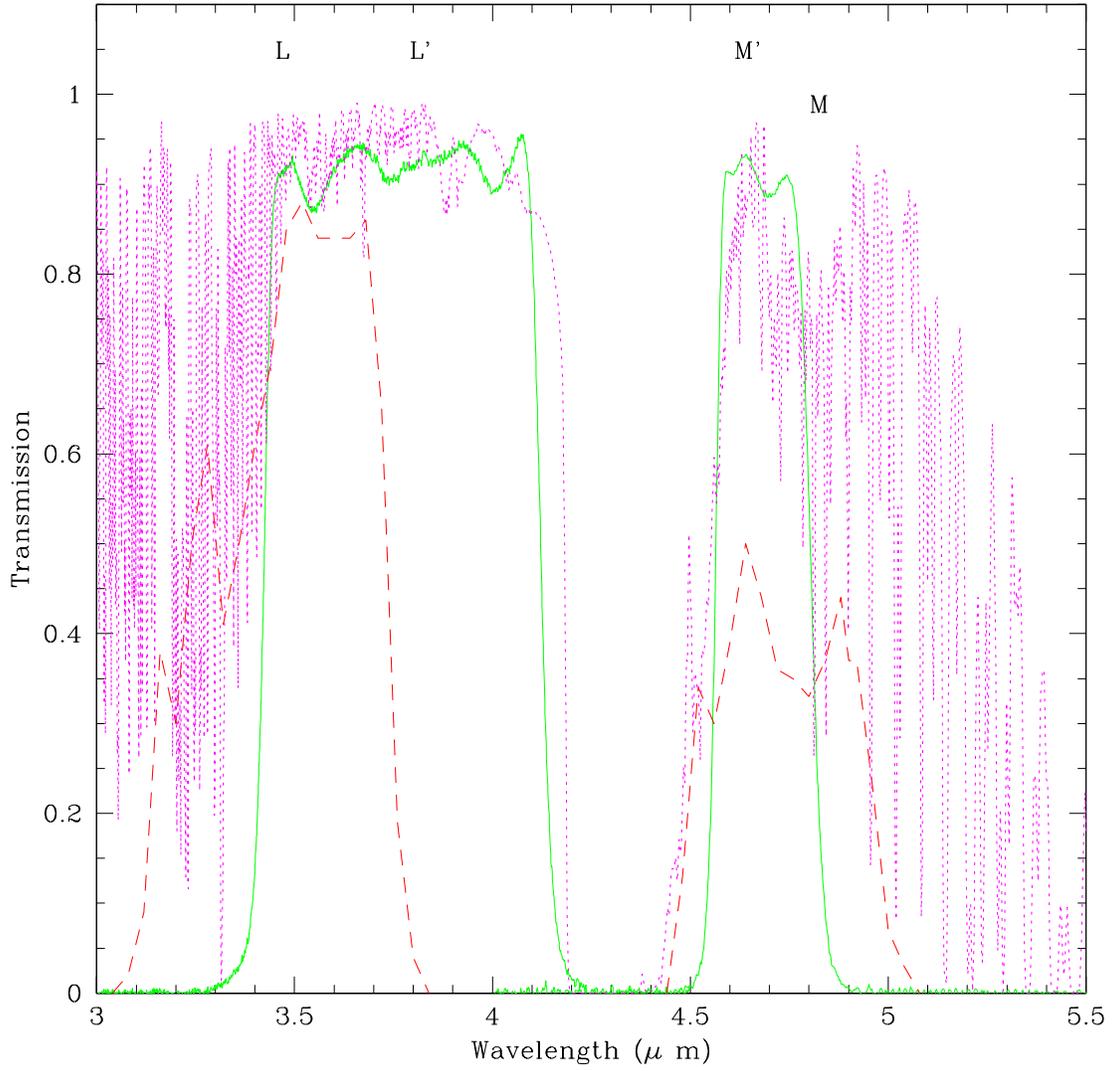}
\caption {The passbands of the traditional Johnson (L, M)
system (dashed lines) and the more recent (L$'$, M$'$) system (solid lines) matched
against the average atmospheric transmission at Mauna Kea Observatory (dotted line).}
\end{figure}

\begin{figure}
\plotone{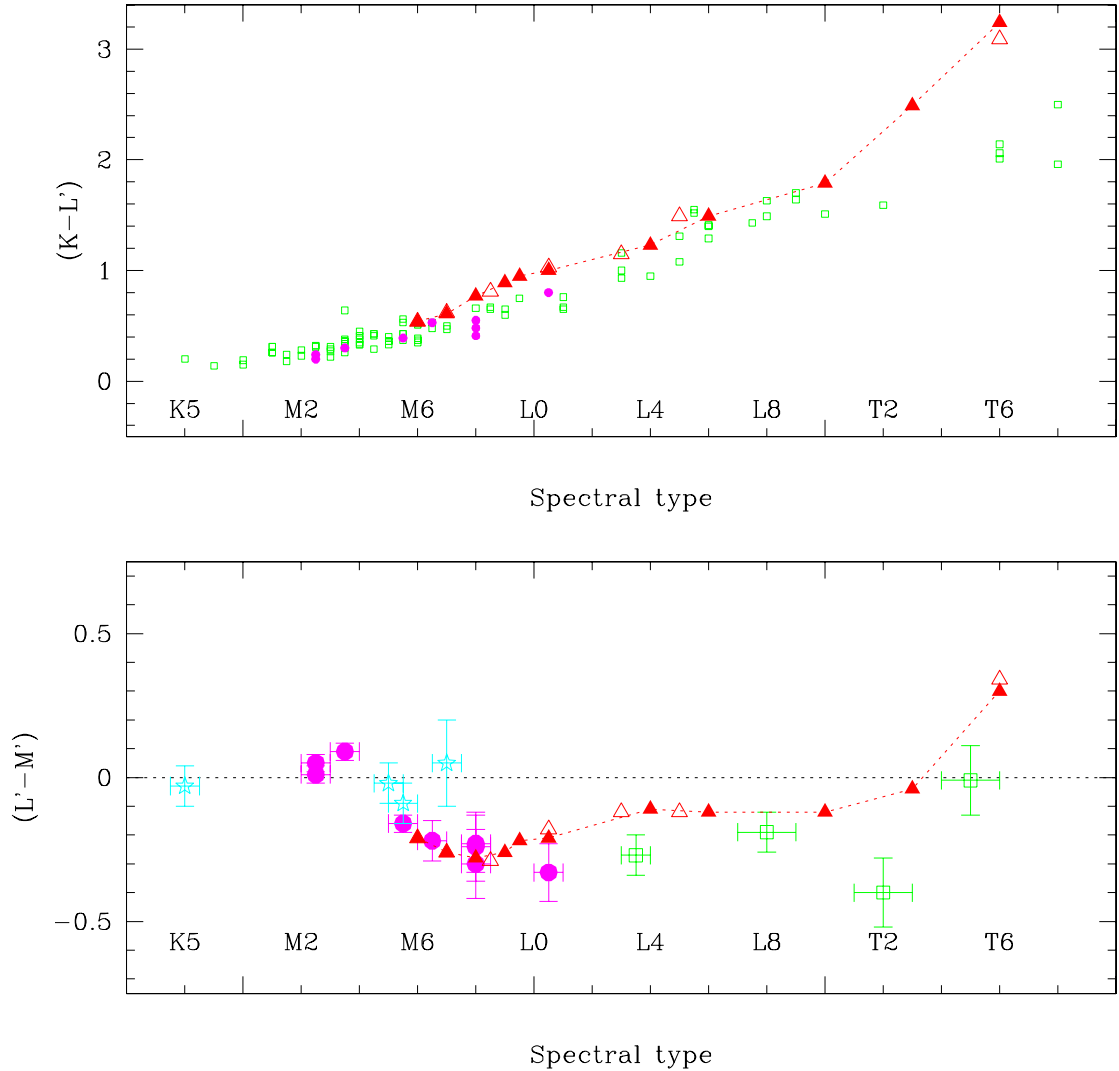}
\caption { (K-L$'$) and (L$'$-M$'$) as a function of spectral type. In the upper panel,
open squares plot data from Leggett (1992), Leggett {\sl et al.} (1998) and
Leggett {\sl et al.} (2001), while
solid points plot the photometry given in Table 2. In the lower panel, data from 
Berriman \& Reid (1987) are plotted as five-point stars, photometry from Leggett {\sl
et al.} (2001) as open squares, and our photometry as solid points. 
We have superimposed predicted colours based on the 
Chabrier {\sl et al.} `DUSTY' models:  solid
triangles, connected by the dotted line, mark data from the 0.1-Gyr isochrone, while
open triangles plot 10-Gyr model predictions. The model colours show little age dependence.
The spectral type/effective temperature conversion and systematics between
the various photometric systems are discussed in the text.  }
\end{figure}

\begin{figure}
\plotone{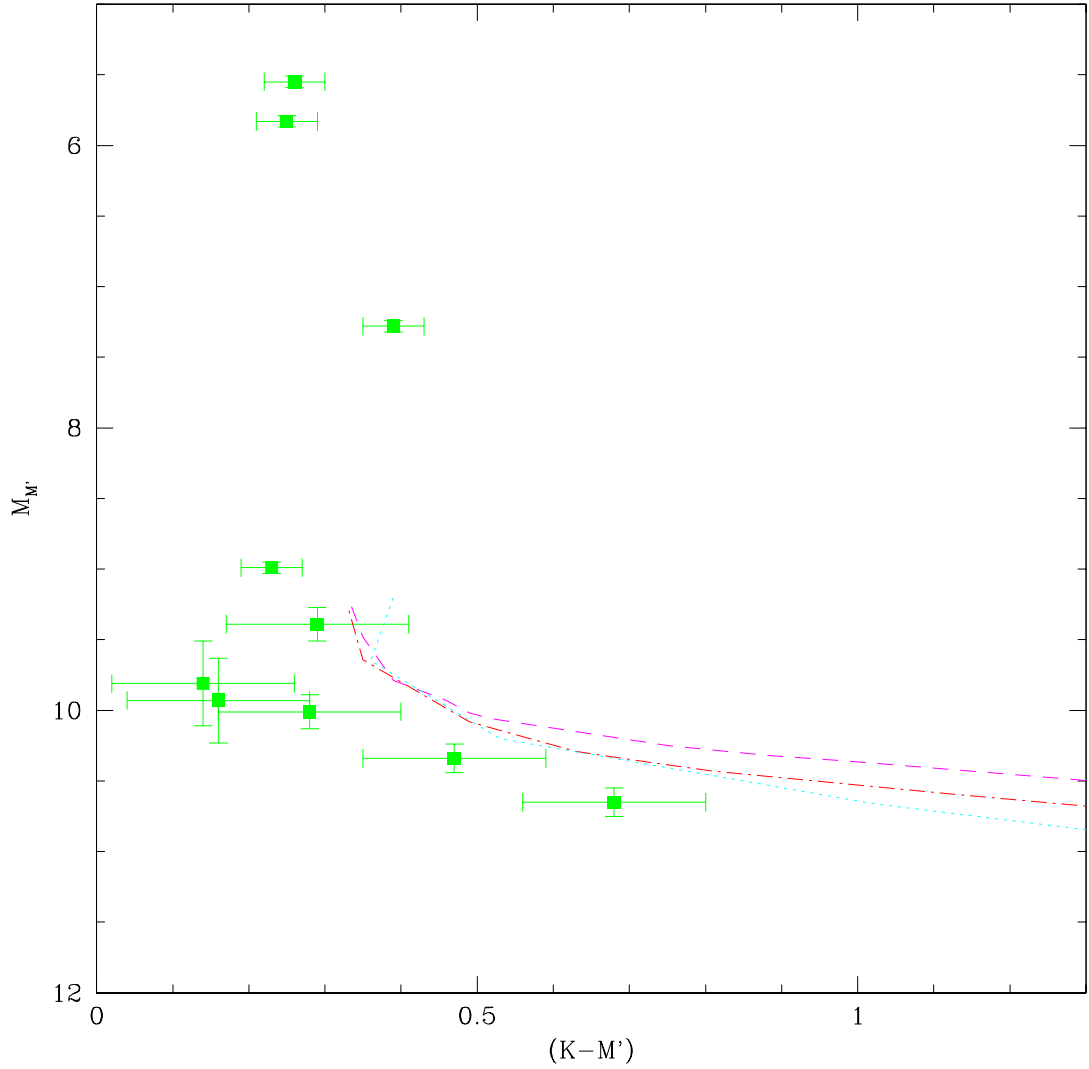}
\caption { A comparison in the (M$_M$, (K-M$'$)) plane between the 0.5-Gyr. 
(dashed), 1-Gyr. (dash-dot) and 5-Gyr (dotted) DUSTY 
isochrone from Chabrier {\sl et al.} (2000) and the observations listed in Table 2.
The faintest data point is for 2M0036+18, from Leggett {\sl et al.} (2001).}
\end{figure}

\end{document}